\documentclass[superscriptaddress,showpacs,preprintnumbers,amsmath,amssymb,aps]{revtex4}
\def\b{\mathbf} \def\d{{\rm d}}\def\U{\Upsilon}

\usepackage{graphicx,dcolumn,bm,color}

\begin{document}

\title{Propulsion by passive filaments and active flagella near boundaries}
\author{Arthur A. Evans}
\email{aevans@physics.ucsd.edu}
\affiliation{Department of Physics}
\author{Eric Lauga}
\email{elauga@ucsd.edu}
\affiliation{Department of Mechanical and Aerospace Engineering, University of California San Diego,  9500 Gilman Drive, La Jolla CA 92093.}
\date{\today}

\begin{abstract}

Confinement and wall effects are known to affect the kinematics and propulsive characteristics of swimming microorganisms.  When a solid body is  dragged through a viscous fluid at constant velocity, the presence of a wall  increases fluid drag, and thus the net  force required to maintain speed has to increase. In contrast, recent optical trapping experiments have revealed that the propulsive force generated by human spermatozoa is decreased by the presence of boundaries.  Here, we use a series of simple models to analytically elucidate the propulsive effects of a solid boundary on passively actuated filaments and  model flagella. For passive flexible filaments actuated periodically at one end, the presence of the wall is shown to
 increase the propulsive forces generated by the filaments  in the case of displacement-driven actuation, while it decreases the force in the case of force-driven actuation.
 In the case of active filaments as models for eukaryotic flagella, we demonstrate that the manner in which a solid wall affects propulsion cannot be known a priori, but is instead a nontrivial function of the flagellum frequency, wavelength, its material characteristics, the manner in which the molecular motors self-organize to produce oscillations (prescribed activity model or self-organized axonemal beating model), and the boundary conditions applied experimentally to the tethered flagellum. In particular, we show that in some cases, the increase in fluid friction induced by the wall can lead to a change in the waveform expressed by the flagella which results in a decrease in their propulsive force. 

\end{abstract}
\pacs{47.63.Gd, 47.15.G-, 87.17.Jj}

\maketitle

\section{Introduction}

In the highly viscous environment inhabited by micro-organisms, locomotion is a difficult task, and one rarely achieved in the absence of fellow organisms, boundaries or other obstacles.  Because drag forces from the fluid dominates inertia, swimming becomes a problem for microscopic life qualitatively different from that of larger organisms such as fish, and nature has evolved several strategies for solving it. Flagellated organisms such as bacteria and spermatozoa utilize the fluid drag anisotropy of slender filaments (flagella) in order to propel themselves through a viscous fluid  \cite{Lauga:2008p421}.  From a biological standpoint, both prokaryotic and eukaryotic flagella serve the same purpose, to propel the organism through the fluid, but from a mechanical standpoint the filaments are quite different. Bacteria such as \textit{E. coli} and \textit{B. subtilis} actuate  passive helical filaments using rotary motors embedded in the cell walls, and whose rotation gives rise to propulsion \cite{Berg:2000p310,bergbook}. In contrast, spermatozoa (and more generally, eukaryotic) flagella are active filaments. They possess  an internal musculature, termed the axoneme, which deforms in a wave-like fashion due to the action of molecular motors. These motors generate time-varying and coordinated bending moments along the flexible flagellum, giving rise to traveling waves,  and propulsion of the cell  \cite{Brennen:1977p30,Lauga:2008p421}.  In that case, the waveform displayed by the cell is a physical balance between the motor activity, the flagellum elasticity, and the fluid forces.

Near solid boundaries, the behavior of both types of swimming cells is strongly affected. Since the governing equations for inertialess fluid flow are time invariant, the geometry of the system fully defines the hydrodynamics.  Bacteria swim in circles near a wall, as the chirality of the flagellar rotation induces a hydrodynamic torque on the body \cite{frymier95,frymier97,Lauga:2006p282}.  Boundaries also tend to hydrodynamically attract swimming cells, and as a result the steady-state distribution of motile cells strongly peaks near walls \cite{rothschild63,winet84a,fauci95,cosson03,woolley03,berke08}. Near walls, large arrays of cilia (short flagella) are known to synchronize,  and display coordinated modes of deformation known as metachronal waves \cite{blake74,Brennen:1977p30,childress81,sleigh88,braybook,salathe07}.

One topic of renewed interest concerns the dynamics of spermatozoa in confinement, as relevant to the situation in mammalian reproduction \cite{fauci06}. Early theoretical studies considered flagellated cells  with waveforms unchanged by the presence of walls. In that case, because of increased drag forces, the cells have to increase their work against the fluid to maintain their waveforms, and as a result they speed up when near a boundary \cite{Reynolds1965,katz74,katz75,katzblake,fauci95}. If instead the cells are assumed to work with a fixed power,  the presence of a boundary leads in general  to a decrease in the swimming speed  \cite{Reynolds1965,katz74}. 

Physically, the speed at which a cell swims is a balance between the propulsive force generated by  its flagellum and the drag from the surrounding fluid. Near a boundary, both the flow field generated by the flagellum and the subsequent propulsion generated are expected to be modified, but in a manner which has not been quantified yet.

Recently, an experimental investigation was carried out using optical trapping on human spermatozoa to investigate the influence of boundaries on force generation. Briefly, spermatozoa cells swimming near and parallel to a cover glass  (distance about 5 $\mu$m) were optically trapped, and then moved to a pre-defined distance of up to 100 $\mu$m from the glass surface.  As the flagellum of the trapped cell continuously beat, the trap power was then gradually attenuated until the cell escaped. The magnitude of the propulsive force applied by the cell, equal to the minimum force required to hold the cell in place by the optical trap,
was found to be decreased by the presence of the glass surface \cite{Lauga10}.  These results indicate that  the cells do not maintain their waveforms, as for a cell with a fixed waveform the propulsive force would increase near the wall. This experimental result suggests therefore that, for eukaryotic cells, the interplay between flagellum elasticity, internal actuation, and hydrodynamics can lead to a non-trivial relationship  between the environment (here, confinement) and the propulsive force generated by the cells.

In this paper, we use a series of simple models to examine the propulsive effects of a solid boundary on passively actuated filaments and  model flagella.  Our work aims at capturing the essential physics that describes the geometric effects on the body, and builds on previous studies of flagellar locomotion far from external influences, both for passive filaments \cite{Wiggins:1998p258,Wiggins:1998p341,Goldstein:1995p188,lauga07_pre,Yu:2006p60}, and for active flagella \cite{Gray:1955p355,Fu:2008p518,Camalet:2000p38,Brokaw71,Brokaw65} (see also Ref.~\cite{Lauga:2008p421} and references therein).  By considering different modeling approaches for the filament actuation, and by quantitatively including the change in viscous friction due to the presence of the wall, we predict analytically the change in flagellar waveform, as well as the resulting change in propulsive force. We demonstrate that the relationship between the wall-flagella distance and the propulsive force it generates is in general non-monotonic. 
For the case of passive flagella actuated at one end, the presence of the wall  increases the propulsive forces generated by the filament dynamics for displacement-driven actuation, but decreases them in the case of force-driven actuation. In contrast, for active filaments we demonstrate that the manner in which a solid wall affects propulsion cannot be known a priori, but is instead a complicated function of the flagella frequency, wavelength, their boundary conditions and the manner in which the molecular motors self-organize to produce oscillations.

The paper is organized as follows. We start with a summary of the general class of elastohydrodynamics problems, and prescribe our course of action for determining the propulsive force and force gradients in the presence of a wall for our models (Sec.~\ref{Elastohydrodynamics}).  Following this we consider a passive filament actuated at one end, and the modification of the thrust it produces in the presence of the no-slip boundary (Sec.~\ref{PassiveFilament}).  We then consider two models for active flagella swimming very close to the wall, first the case of a prescribed internal sliding force, and then the  more realistic case of flagellar beating via self-organization of molecular motors in the axoneme (Sec.~\ref{ActiveFlagellum}). In both cases, we determine the propulsive force, and how it is modified by the presence of the wall. We finish with a discussion of our results in the context of spermatozoa locomotion (Sec.~\ref{discussion})

\section{Elastohydrodynamics and setup}
\label{Elastohydrodynamics}

\begin{figure}[t]
\includegraphics[width=.6\textwidth]{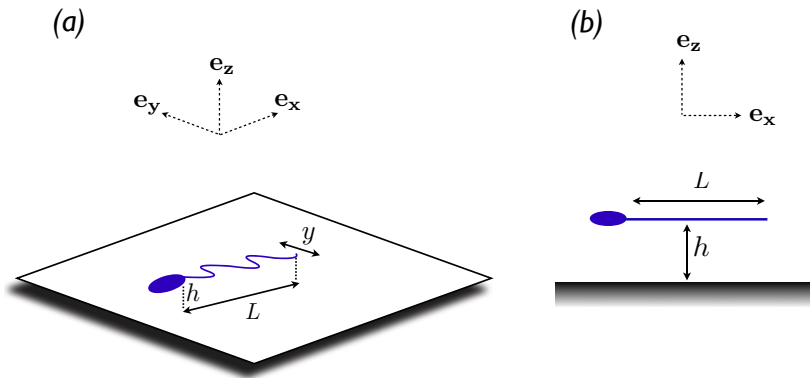}
\caption{\label{schematic} (color online) Schematic diagram of a flagellated cell (or passive flexible filament) in the presence of a stationary no-slip boundary.  
(a): $3/4$ view of the model system.  The planar beating of the flagellum or filament takes place parallel to the plane of the wall ($x,y$), and the cell is located at a distance $h$ from the surface. For simplicity we ignore the head effects in this work; this is a reasonable assumption as many experiments to measure the propulsive force would anchor the head in place (e.g. an optical trap or micropipet). (b): Side-view of the model system.  Note that from this angle the flagellum or filament appears only as a straight rod, as it is assumed to beat along the $y$ direction.}
\end{figure}

\subsection{Setup}

The physical system that we will investigate is illustrated schematically in Fig.~\ref{schematic}. We consider a single flagellated cell (or a synthetic device with a flagellum-like filament), for which the flagellum undergoes planar beating a distance $h$ from the surface of a solid boundary. The plane of actuation is assumed to remain parallel to the plane of the wall (directions $x$ and $y$ in Fig.~\ref{schematic}).  
In the experiment of Ref.~\cite{Lauga10}, when the cells are trapped, their plane of beating is parallel to the surface, and thus it is a reasonable approximation to assume that it remains so at different heights. We will consider both  cases of a passive filament actuated at one end (Sec.~\ref{PassiveFilament}) as well as an active filament with internally distributed actuation (Sec.~\ref{ActiveFlagellum})

\subsection{Hydrodynamics}

Since the Stokes equations are time-invariant, the geometry of the system completely defines the fluid dynamics at zero Reynolds number.  Although the governing equations are themselves linear, nonlinearities in the shapes of swimming organisms can make calculating the flow difficult.  Furthermore, because the flow is determined by the instantaneous shapes of the surfaces immersed in the flow, nonlocal hydrodynamic effects can make the calculations impossible to do  without numerical analysis.

Fortunately, for flagellated organisms the slenderness of the swimming surface in question leads to several simplifications that can be made to the fluid dynamics.  Using the asymptotic limit of a slender filament, the force acted by the fluid on the moving filament  is approximately given by \cite{Cox:2006p33}
\begin{gather}\label{fluid}
 \b{f}_{fl}=-\left[\zeta_{\|}\b{\hat{t}\hat{t}}+\zeta_{\bot}(\b{I}-\b{\hat{t}\hat{t}})\right]\cdot\b{v},
 \end{gather}
where $\zeta_{\bot}$ and $\zeta_{\|}$   are the force-velocity resistance coefficients derived from resistive force theory $(\zeta_{\|}< \zeta_{\bot})$ \cite{Lauga:2008p421},  and accounting for filament motion perpendicular and parallel to the filament axis respectively; here $\b{\hat{t}}$ is the unit tangent vector along the filament, $\b{I}$ is the identity tensor, and $\b{v}$ is the velocity of the filament as it moves through the quiescent fluid.

In addition to slenderness, for small amplitudes of the flagellar beat the geometric terms simplify considerably.  Although real flagella beat with a large amplitude, experiments and numerics have shown that corrections to linearized dynamical shape equations are sub-leading \cite{RiedelKruse:2007p549,Hilfinger:2009p539,yu06}.  In the linearized regime, only the normal velocity component of the filament is important for calculating the fluid force, as the tangential motion enters only for higher-order curvature or  bending of the filament.  In that case, we can represent the amplitude of the filament perpendicular to the propulsive direction as a function $y(x,t)$, and  the fluid drag on the filament can be simply expressed in terms of the amplitude $y$ as $\b{f}_{fl}\approx -\zeta_{\bot} \partial y/\partial t\, {\bf e}_y$.

\subsection{Flexibility and activity}

In order to model  propulsion, we need to balance the expression for the fluid force on the filament with the internal forces,   $\b{f}_{int}$, of the model flagellum.  In the case of a passive filament actuated at one end, only the elastic forces contribute to this term.  For a model of spermatozoa, the internal distribution of molecular motors in the axoneme leads to  an active bending moment that contributes to force and torque balance.  In the inertialess realm of low Reynolds number that is inhabited by the swimming cells that we examine, the total force on each infinitesimal element of the flagellum sums to zero, and therefore mechanical equilibrium is written as 
\begin{equation}
\label{balance}
\b{f}_{fl}+\b{f}_{int}=0.
\end{equation}

Thin passive filaments are well modeled by the elastic beam theory \cite{landau_lifshitz_elas}. Their  elastic strain energy associated with deformation is given by 
\begin{gather}
E_{el}=\frac{A}{2}\int_0^L{\kappa(s)^2ds},
\end{gather}
where $A$ is the bending rigidity of the filament, $L$ its length, and $\kappa(s)$ its local curvature along the arclength, $s$.

\begin{figure}[t]
\includegraphics[width=.4\textwidth]{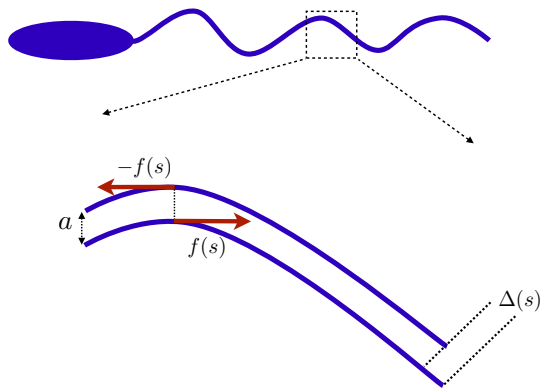}
\caption{\label{schematic2} (color online) Schematic of a flagellated micro-organism swimming via an active flagellum.  The inset shows the model of the active filament that we use in this analysis (adapted from  Ref.~\cite{Camalet:2000p38}).  The two pieces of the filament slide past one another, each exerting an equal and opposite active force $f(s)$ that is dependent on the position along the flagellum (the arclength $s$).  This produces a sliding displacement $\Delta$. In the simplest case, that of purely elastic response, and for small amplitude motion, this displacement $\Delta$ induces a restoring force that is proportional to the magnitude of $f(s)$.  The distance between filaments, a, has been exaggerated for viewing purposes.}
\end{figure}

For an active filament that is powered by an internal musculature, such as a eukaryotic flagellum, we must consider not only similar elastic restoring forces, but also any internal actuation forces.  We use in this paper the model of Camalet \& Julicher  \cite{Camalet:2000p38} to address the mechanics of active filaments, as illustrated in  Fig.~\ref{schematic2}. The active filament is assumed to be composed of two inextensible elastic beams which are attached to the basal (or head) region, but allowed to bend relative to one another, and acted upon by a distribution of equal and opposite active force $f(s)$.
As the filaments bend, they induce a distribution of sliding displacements, $\Delta(s)$,  given geometrically by 
\begin{gather}
\Delta(s)=\int_0^s{a\kappa(s')ds'},
\end{gather}
where $a$ is the fixed distance between the filaments at the base.  The work done by the filament against the internal forces is then added to the enthalpy functional, and we get 
\begin{gather}
\label{enthalpy}
E=\int_0^L{ds\left[\frac{A}{2}\kappa^2(s)+f\Delta(s)\right]}.
\end{gather} 
In the case of a passively actuated filament, the internal forces are zero, and thus only the elastic contributions enter the equations of motion.  Extremizing the energy given by Eq.~\eqref{enthalpy}, for a particular form of the active forces, yields the total internal force per length on the flagellum, $\b{f}_{int}$, which must then be equal and opposite to the fluid force.

\subsection{Propulsion and wall effects}

Given a system, either the passive or active filament, and a set of boundary conditions for the flagellum corresponding to a given physical situation, the elastohydrodynamics balance allows us to solve for the flagellar beating pattern and thus for the propulsive force.
For small-amplitude motion, the propulsive force, defined as the force acted by the beating flagellum on the  surrounding fluid when it is beating but not swimming,  is given by
\begin{gather}\label{propulsion}
\textbf{F}=\left[(\zeta_{\|}-\zeta_{\bot})\int_0^L{\frac{\partial y}{\partial t}\frac{\partial y}{\partial x}dx}\right] \textbf{e}_x,
\end{gather}
where  $\b{e}_x$ is the unit vector along the average position of the beating filament   \cite{lauga07_pre} (see Fig.~\ref{schematic}). In particular, if $y(x,t)$ deforms as a pure traveling wave propagating in the $+x$ direction, $y(x,t)=y_0(x-ct)$, we get 
$\textbf{F}=\left[(\zeta_{\bot}-\zeta_{\|})c\int_0^L (\partial y_0/\partial x)^2dx\,\right] \textbf{e}_x$. In that case, the force on the fluid is in the $+x$ direction, and therefore the force on the filament, and the swimming direction if it was free to swim, is in the $-x$ direction.

To determine the values of $\zeta_{\bot}$ and $\zeta_{\|}$ in Eq.~\eqref{propulsion}, we look at previous work calculating friction coefficients for slender bodies near walls \cite{Brennen:1977p30}.  For a slender body of radius $a$ and length $L$ in an unbounded fluid the drag coefficients can be calculated asymptotically in the slender limit $a\ll L$, and are given by
\begin{gather}\label{inf}
(\zeta_{\bot})_{\infty}\approx\frac{4\pi\mu}{ln({2L}/{a})+C_1}\quad(\zeta_{\|})_{\infty}\approx\frac{2\pi\mu}{ln({2L}/{a})+C_2}
\end{gather}
where $\mu$ is the fluid viscosity and  $C_1$ and $C_2$ are $O(1)$ constants that depend on the specific geometry of the filament.  Near a boundary, if $h$ is the  distance between the filament and  the wall, the situation  relevant to the experiments in Ref.~\cite{Lauga10} is that of $h \lesssim L $. In this near-field limit, the resistance coefficients relevant to the  planar beating geometry  considered in Fig.~\ref{schematic} are given by 
\begin{gather}
\label{resist}
\zeta_{\bot}\approx\frac{4\pi\mu}{\ln({2h}/{a})},\quad \zeta_{\|}\approx \frac{1}{2}\zeta_{\bot}.
\end{gather}
The far-field limit, $h \gtrsim L $, yields only a small correction to the values of  $\zeta_{\bot}$ and $\zeta_{\|}$ in Eq.~\eqref{inf} which would likely be too small to be measured experimentally, and thus we will not consider this case here. For a comprehensive account of these calculations see Ref.~\cite{Brennen:1977p30} and references therein.  Experiments on sedimenting cylinders near boundaries were considered in Ref.~\cite{Stalnaker79}, where
it is shown that  the difference in drag between a cylinder far from a wall, and one in the near-field regime that we consider can easily be as much as $50\%$. In particular, the result of Eq.~\eqref{resist} implies that there is a gradient in the fluid friction if the body changes its distance, $h$,  from the wall. The friction gradient is given by
\begin{gather}\label{zeta_wall}
\frac{\d\zeta_{\bot}}{\d h}=-\frac{\zeta_{\bot}^2}{4h\pi\mu},
\end{gather}
which is always negative, reflecting physically that viscous forces are increased by the presence of a solid boundary.

For a slender rod dragged through a viscous fluid with a fixed velocity, the sign of this force gradient implies that moving the rod closer to the  wall will increase the force required to maintain its speed.   For an actuated filament, friction from the surrounding fluid plays a dual role. It  first affects the propulsive force, as given by Eq.~\eqref{propulsion}, through change in the drag coefficients. In addition, for either boundary- or internally-actuated filaments, the change in  $\zeta_{\bot}$ and $\zeta_{\|}$ in Eq.~\eqref{fluid} modifies the force balance, Eq.~\eqref{balance}, thereby changing the shape of the flagellum, and therefore affecting the propulsion in Eq.~\eqref{propulsion} through a modification of $y(x,t)$. It is this interplay between viscous friction, elasticity, and  activity that we propose to quantitatively analyze in this paper. As we detail below, it usually results in a non-trivial and non-monotonic relationship between wall distance and propulsive force.

%%%%%%%%%%%%%%%%%%%%%%%%%%%%
%%%%%%%%%%%%%%%%%%%%%%%%%%%%
%%%%%%%%%%%%%%%%%%%%%%%%%%%%
%%%%%%%%%%%%%%%%%%%%%%%%%%%%

\section{Passive filaments}
\label{PassiveFilament}

We first consider the case of a passive elastic filament driven at one end. In that case, the  elastic energy is
\begin{gather}
\label{elastic_passive}
E_{el}=\int_0^L{\frac{A}{2}\left(\frac{\partial^2y}{\partial x^2}\right)^2dx},
\end{gather}
where the linearized regime allows for the approximation of the curvature $\kappa$ by the concavity of the function $y(x)$.  
Calculating the  functional derivative of the energy  in Eq.~(\ref{elastic_passive}) leads to the elastic force density which, when balanced with the  fluid force density, results in the  linearized dynamics equation 
\begin{gather}
\label{passive_eqn}
\zeta_{\bot}\frac{\partial y}{\partial t}=-A\frac{\partial^4y}{\partial x^4},
\end{gather}
as obtained in previous studies \cite{Wiggins:1998p258,lauga07_pre}. We will consider a harmonic driving  with  frequency $\omega$, and by focusing only on post-transient effects, we will assume a similar periodic dynamics for the filament.

Since the filament is driven only at one end, the boundary conditions at the tail ($x=L$) end are
\begin{eqnarray}
A\frac{\partial^2y}{\partial x^2}(L,t)&=&0,\\
A\frac{\partial^3y}{\partial x^3}(L,t)&=&0.
\end{eqnarray}
This guarantees that the ends of the filament are force- and torque-free.  At the driving position ($x=0$), different types of physical actuation could be implemented experimentally, leading to  different boundary conditions.  If the tangent angle is prescribed, with the position of the filament fixed in place, the boundary conditions are
\begin{eqnarray}
y(0,t)&=&0, \\
\frac{\partial y}{\partial x}(0,t)&=&\epsilon \cos\omega t ,
\end{eqnarray}
where $\epsilon\ll1$ is the magnitude of the tangent angle deviation from horizontal and $\omega$ the driving frequency (pivoted case).   Alternatively, for experiments that involve manipulating a passive filament via optical tweezers, the driving position  would be made to oscillate harmonically under torqueless conditions, leading to the boundary conditions
\begin{eqnarray}
y(0,t)&=&y_0 \cos\omega t ,\\
A\frac{\partial^2 y}{\partial x^2}(0,t)&=&0,
\end{eqnarray}
where $y_0$ is the magnitude of the oscillation in space (tweezed case). Finally, we could also consider the case where an oscillating force or torque is applied to the driving end leading to boundary conditions
\begin{eqnarray}
y(0,t)&=&0, \\
A\frac{\partial^2 y}{\partial x^2}(0,t)&=&M_0 \cos\omega t \,\,\,{\rm (Torqued),} \\
A\frac{\partial^3 y}{\partial x^3}(0,t)&=&F_0 \cos\omega t \,\,\,{\rm (Forced).}
\end{eqnarray}
Here $M_0$ and $f_0$ are the magnitudes of the time-varying torque and force, respectively. These boundary conditions described above are experimentally realizable, for example using micro-pipets.

Since we ignore transient effects, the steady solution for the amplitude can be written as $y(x,t)=\Re\left\{\tilde{y}(x)e^{-i\omega t}\right\}$.  Additionally, we can define a natural length scale through the dimensionless ``sperm number", $Sp=L/\ell_{\omega}$ where $\ell_{\omega}=(A/\zeta_{\bot}\omega)^{1/4}$; $Sp$ is the only dimensionless number in this problem, and as such it fully governs the filament dynamics. For $Sp\ll1$, the �penetration length� $\ell_\omega$ is much larger than the length of the filament, and thus oscillations will not decay along the length of the flagellum.  For a passive filament, this is equivalent to a very rigid rod being actuated back and forth.  Conversely, for $Sp\gg1$, oscillations decay very quickly, indicating a floppy string or very viscous fluid.  For an active filament, there are additional length scales that must be considered, as  will be discussed later.

We nondimensionalize time by $1/\omega$, the direction $x$ along the filament by $L$, and the filament amplitude by $y_0$ which is  specified in the tweezed case, and can be related to the 
forcing parameters in the other three cases. Specifically we have $y_0 =\epsilon L$ for pivoted actuation, $y_0 = F_0L^3/A$ for forced actuation, and $y_0 = M_0L^2/A$ for the torqued condition.  By doing so, we obtain the following dimensionless equation for the amplitude $\tilde{y}$ as
\begin{gather}
\label{amplitude}
\tilde{y}''''-iSp^4 \tilde{y}=0.
\end{gather}
The dimensionless boundary conditions for the tail  are thus given by
\begin{eqnarray}
\label{bc_1}\tilde{y}''(1)&=&0, \\
\label{bc_2} \tilde{y}'''(1)&=&0,
\end{eqnarray}
while the various possible boundary actuations at the driving end are
\begin{eqnarray}
\nonumber \tilde{y}(0)&=&0,\,\,\,\,\, \tilde{y}'(0)=1\,\,\,\, {\rm (Pivoted),} \\
\nonumber \tilde{y}(0)&=&1,\,\,\,\,\, \tilde{y}''(0)=0\,\,\,\, {\rm (Tweezed),} \\
\nonumber \tilde{y}(0)&=&0,\,\,\,\,\, \tilde{y}'''(0)=1\,\,\,\, {\rm (Forced),} \\
\nonumber \tilde{y}(0)&=&0,\,\,\,\,\, \tilde{y}''(0)=1\,\,\,\, {\rm (Torqued),}
\end{eqnarray}

With a solution to the amplitude equation,  Eq.~(\ref{amplitude}), we find the total force exerted on the filament by the fluid.  Because we consider harmonic actuation, we only look at the time-averaged propulsive force, which is given, in a dimensional form, by
\begin{gather}\label{force_passive}
\langle F \rangle=\frac{1}{4}\frac{\omega\zeta_{\bot}y_0^2I(Sp)}{Sp^4},
\end{gather}
where $I(Sp)$ is a dimensionless integral  defined by
\begin{gather}
I(Sp)=Re\left[\frac{1}{2}\tilde{y}_{xx}\tilde{y}^*_{xx}-\tilde{y}_x\tilde{y}^*_{xxx}\right]
(x=0),
\end{gather}
where an asterisk denotes the complex conjugate, and $Re $ the real part.  This expression is a direct consequence of the integral for the propulsive force, Eq.~\eqref{propulsion}, being a total derivative when the filament is passive, and thus the function depends only on the non-dimensional amplitude $\tilde{y}$ evaluated at the endpoints \cite{Wiggins:1998p258}; the free end $x=1$  does not contributed because of Eqs.~\eqref{bc_1}-\eqref{bc_2}.

If we then define a scaling function $Z(Sp)=I(Sp)/Sp^4$ to contain all of the dependence on $Sp$ for the propulsive force, we have that the four cases are given by
\begin{gather}
\langle F \rangle=\frac{1}{4}\omega\zeta_{\bot}\epsilon^2L^2Z(Sp)\,\,\,\, {\rm (Pivoted),} \\
\langle F \rangle=\frac{1}{4}\omega\zeta_{\bot}y_0^2Z(Sp)\,\,\,\, {\rm (Tweezed),} \\
\langle F \rangle=\frac{1}{4}\frac{\omega\zeta_{\bot}F_0^2L^6}{A^2}Z(Sp)\,\,\,\, {\rm (Forced),} \\
\langle F \rangle=\frac{1}{4}\frac{\omega\zeta_{\bot}M_0^2L^4}{A^2}Z(Sp)\,\,\,\, {\rm (Torqued),}
\end{gather}
and each of the scaling functions is unique to the four boundary conditions that solve the amplitude Eq.~(\ref{amplitude})

We can now examine how the propulsive force changes as the filament changes its distance $h$ to the  wall.  Using the chain rule, the force gradient is given by
\begin{eqnarray}
\frac{\d}{\d h} \langle F \rangle
\label{first}&=&\frac{1}{4}\omega y_0^2 \left(\frac{\d\zeta_{\bot}}{\d h}Z+\zeta_{\bot}\frac{\partial Z}{\partial h}\right)\\
&=&\frac{1}{4}\omega y_0^2 \left(\frac{\d\zeta_{\bot}}{\d h}Z+\zeta_{\bot}\frac{\partial Z}{\partial Sp}\frac{\partial Sp}{\partial \zeta_{\bot}}\frac{\d \zeta_{\bot}}{\d h}\right)\\
\label{grad}
&=&\frac{1}{4}\omega y_0^2\frac{\d\zeta_{\bot}}{\d h}\left(Z +\frac{1}{4}Sp\frac{\partial Z}{\partial Sp} \right),
\end{eqnarray}
where we have used the definition of $Sp$ to take a partial derivative with respect to the resistive coefficient.  Since $Sp$ is monotonic in the resistive coefficient $\zeta_{\bot}$, and from Eq.~\ref{resist}, we see that $\zeta_{\bot}$ itself is a monotonic function of h, we can examine the qualitative behavior of the force gradient by recasting the derivatives with respect to $h$ in terms of the dimensionless number $Sp$.  
It is important to note at this point that the  variation of $Sp$ with the distance between the flagellum and the wall is weak, with a scaling $Sp \sim [\log h/a]^{-1/4}$, and thus even large changes in the distance from the wall will produce small changes in $Sp$.  Quantitative details are discussed in Sec.~\ref{prescribed}, and for a bull spermatozoa cell
with $Sp\approx 7$ away from a boundary, bringing the cell closer to the wall will lead to a typical increase in $Sp$ of about 10\%.

The force gradient is now composed of two terms: the first is due entirely to the changing fluid friction, but the second is a more complicated effect that incorporates the elasticity of the filament through the shape change.  This term reflects the fact that, in the presence of a wall, the elastohydrodynamic penetration length is a function of the distance from the wall. Changing that distance changes the essential character of viscous-induced oscillation in the filament.  Due to the competition between these terms, it is now not necessarily the case that this gradient be negative, as we would expect for the case of the rigid rod.  
From a dimensional standpoint, we  can write
\begin{equation}
\frac{\d}{\d h} \langle F \rangle  = \frac{1}{4}\omega y_0^2 \bigg|\frac{\d\zeta_{\bot}}{\d h}\bigg|  Z'
\end{equation}
where we have therefore, since ${\d\zeta_{\bot}}/{\d h}$ is negative (see Eq.~\ref{zeta_wall}),
\begin{equation}
Z'=-\left(Z +\frac{1}{4}Sp\frac{\partial Z}{\partial Sp} \right)\cdot
\end{equation}

\begin{figure*}[t]
\includegraphics[width=.8\textwidth]{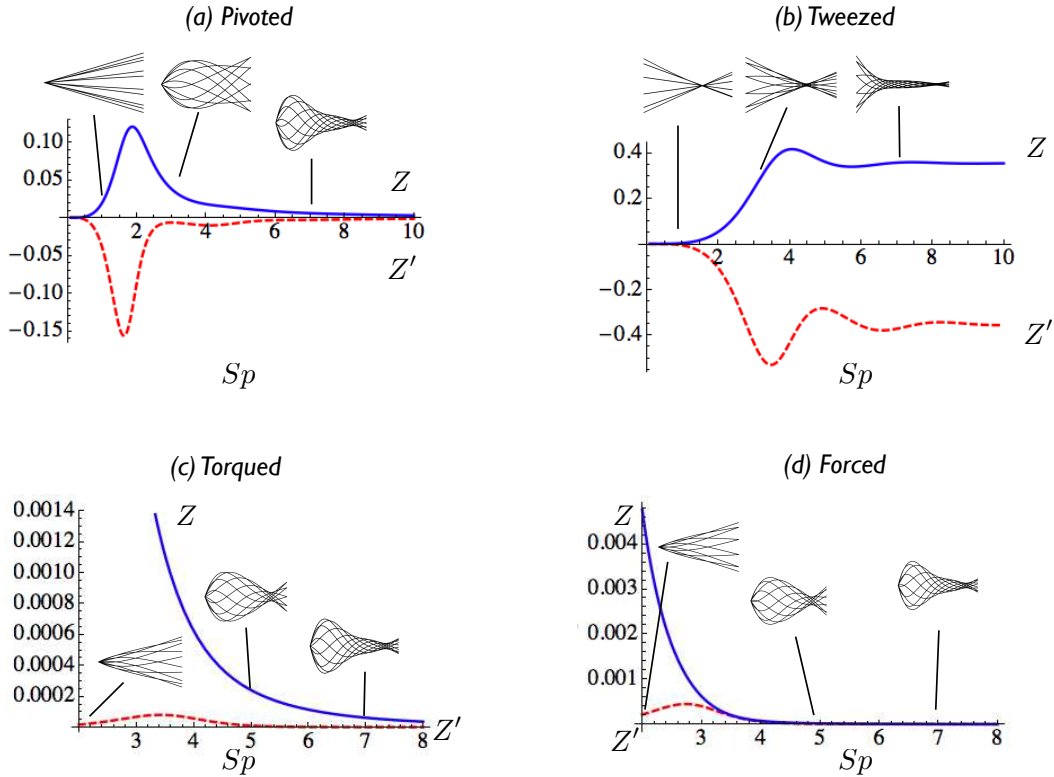}
\caption{\label{passive} (color online) Passive filament: Normalized propulsive force, $Z$, (blue solid line /top curve) and propulsive force gradient, $Z'$ (red dashed line/bottom curve) as a function of the dimensionless parameter $Sp$, for all four boundary conditions.  Shapes of the filament for different $Sp$ are superimposed.  
(a):  Pivoted case; 
(b): Tweezed actuation;   
(c): Torqued condition;
(d):  Forced boundary condition.
In the case of displacement-driven actuation, $\d Z / \d h$ is negative; since $\zeta_\perp$ is also a decreasing function of $h$ (see Eq.~\ref{zeta_wall}), the time-average propulsive force, Eq.~\eqref{force_passive}, is a decreasing function of $h$, and  the presence of a boundary always increase the propulsion of the filament.  On the contrary, for force-driven actuation, the sign of $\d Z / \d h$ is positive, indicating that the presence of the boundary decreases the propulsive force.}
\end{figure*}

To represent the change of the force with the wall distance, we plot in Fig.~\ref{passive} the dimensionless force, $Z$  (top, blue solid line) and force gradient, $Z'$ (bottom, red dashed line) for all four different boundary conditions.   Representative shapes of the filament over an entire period of oscillation are also shown for various values of $Sp$.  For low values of $Sp$, the filament behaves like a rigid rod, while for larger values the decay length of the actuated filament becomes apparent.  In both the pivoted and the tweezed cases, there is a maximum propulsive force (blue curve) that occurs for $Sp>1$: $Sp\approx 4$ for the tweezed case, $Sp\approx 2$ for the pivoted case.  For the torqued and forced cases, the maxima occur for $Sp<1$, and as $Sp$ approaches zero these conditions leave the linearized regime, and the model breaks down.

As can be seen in Fig.~\ref{passive}, $Z$ is always positive meaning that the actuation at $x=0$ leads to filament dynamics pushing against the fluid always in the $+x$ direction. In addition, we see that $Z'$ is always negative for the pivoted and tweezed cases ({displacement-driven} actuation), indicating that the force (or $Z$) is a decreasing function of the distance to the wall, $h$.  The time-average propulsive force, Eq.~\eqref{force_passive}, generated is therefore always increased by the presence of a boundary.  As such, it is somewhat similar to the increase in the drag on a body driven at a constant velocity by the presence of a boundary.  Note  however that $Z'$ displays a non-monotonic dependence on $Sp$ (and therefore on $h$, since $Sp$  is a monotonically decreasing function of $h$), which is due entirely to the change in the filament shape accompanying the  change in height. In contrast, in the torqued and forced cases ({force-driven} actuation), we see that  $Z'$ is positive, indicating that the opposite is true, and the propulsive force is now decreased by the presence of the wall. These results are reminiscent of   
early work showing a similar contrast between two-dimensional swimming with fixed kinematics or fixed hydrodynamic power  \cite{Reynolds1965,katz74,katz75,katzblake}.

%%%%%%%%%%%%%%%%%%%%%%%%
%%%%%%%%%%%%%%%%%%%%%%%%
%%%%%%%%%%%%%%%%%%%%%%%%
%%%%%%%%%%%%%%%%%%%%%%%%
%%%%%%%%%%%%%%%%%%%%%%%%

\section{Active flagella}
\label{ActiveFlagellum}

We now turn to the case of  an active filament as a model for a eukaryotic flagellum, and investigate how  the  changes in  hydrodynamic drag induced by the wall and the response of the flagellar amplitude couples to the internal activity.  As before, we examine force balance, but for an active flagellum we retain the internal forcing term, and the linearized dynamics equation becomes  \cite{Camalet:2000p38,Fu:2008p518,Hilfinger:2009p539}
\begin{gather}
\label{eom}
\zeta_{\bot}\frac{\partial y}{\partial t}=-A\frac{\partial^4y}{\partial x^4}+a\frac{\partial f}{\partial x}\cdot
\end{gather}

Since the total force and moment on the filament must vanish, this imposes boundary conditions on the distal, or ``tail", end as
\begin{eqnarray}
-A\frac{\partial^3 y}{\partial x^3}(L,t)+af(L)&=&0, \\
-A\frac{\partial^2 y}{\partial x^2}(L,t)&=&0.
\end{eqnarray}

Regarding the other side  of the flagellum (its ``head'' side), we will  examine two different types of boundary conditions, either clamped or hinged.  For a clamped filament, physically corresponding to a cell that is immobilized via micro-pipet, both the  head and its tangent angle cannot change, and
 the boundary conditions are thus
\begin{eqnarray}
 y(0,t)&=&0, \\
\frac{\partial y}{\partial x}(0,t)&=&0.
\end{eqnarray}
In the hinged case, the head remains fixed, but the tangent angle can move torquelessly. This is the situation that would take place in the presence of an optical trap, so the boundary conditions become
 \begin{eqnarray}
 y(0,t)&=&0, \\
 A\frac{\partial^2 y}{\partial x^2}(0,t)+a\int_0^L{f(x)}dx&=&0.
 \end{eqnarray}

If we non-dimensionalize $x$ by $L$ and time by $1/\omega$ as in the previous section, but additionally scale the magnitude of the flagellar beat $y_0$ by $af_0L^3/A$, where $f_0$ is the magnitude of the internal force, then the equation for mechanical equilibrium becomes
\begin{gather}
\tilde{y}''''-iSp^4 \tilde{y}=\frac{\partial \tilde{f}}{\partial x}\cdot
\end{gather}
The dimensionless version of the boundary conditions are
\begin{eqnarray}
-\tilde{y}'''(1)+\tilde{f}(1)&=&0,\\
\tilde{y}''(1)&=&0,
\end{eqnarray}
for the tail end, and $\tilde{y}(0)=0$ and one of the following for the head
\begin{eqnarray}
\tilde{y}'(0)&=&0 \,\,\,{\rm (Clamped)}, \\
\tilde{y}''(0)+\int_0^1{\tilde{f}(x)dx}&=&0 \,\,\,{\rm (Hinged).}
\end{eqnarray}

In order to consider the response of the flagellum due to changing the distance from a wall, and thus changing the friction of the fluid, the mechanism of the axoneme itself must be taken into account, {\it i.e.}, a model for $f$ must be prescribed.  We proceed in the next two sections by considering two such models.
 
%%%%%%%%%%%%%%%%%%%%
%%%%%%%%%%%%%%%%%%%%

\subsection{Prescribed activity}
\label{prescribed}

In the first approach, we consider that the active force per unit length takes the form of a prescribed traveling wave, i.e. $f(x,t)=\Re\{f_0e^{ikx-i\omega t}\}$ \cite{Fu:2008p518}, whose frequency ($\omega$) and wavenumber ($k$) are not modified by the presence of a boundary.  This enables us to completely specify the filament shape, its propulsive force, and the propulsive force gradient, with two dimensionless numbers: $Sp$ and $kL$.   The dimensionless equation of mechanical equilibrium is now written as
\begin{gather}
\tilde{y}''''-iSp^4 \tilde{y}=ikLe^{ikLx},
\end{gather}
with boundary conditions 
\begin{eqnarray}
\tilde{y}(0)&=&0, \\
\tilde{y}'''(1)&=&e^{ikL} ,\\
\tilde{y}''(1)&=&0,
\end{eqnarray}
and either of the following
\begin{eqnarray}
 \tilde{y}'(0)&=&0\,\,\,\,\ {\rm (Clamped),} \\
 \tilde{y}''(0)&=&-\frac{(1-e^{ikL})}{ikL} \,\,\,\, {\rm (Hinged).}
\end{eqnarray}

It is then straightforward to calculate the propulsive force on the fluid due to the active filament, which is given by
\begin{gather}
\label{activeforce}
\langle F \rangle =\frac{1}{4} \omega \zeta_{\bot} y_0^2\Upsilon(Sp,kL),
\end{gather}
where the scaling function $\Upsilon(Sp,kL)$ is defined as the dimensionless integral
\begin{gather}
\Upsilon(Sp,kL)=Im\left[\int_0^1{\tilde{y}^*\frac{\partial \tilde{y}}{\partial x}dx}\right]\cdot
\end{gather}
Here  $Im $ denotes the imaginary part.  Note that the amplitude of the flagellar beat is proportional to the magnitude of the active force, and thus  the propulsive force scales quadratically with it.  This dependence has been scaled out of the propulsive force, and is assumed to remain constant.

As in the previous section, the force gradient can be calculated using the chain rule and we get
\begin{eqnarray}
\frac{d}{dh}\langle F \rangle &=& \frac{1}{4}\omega y_0^2 \left(\frac{\d \zeta_{\bot}}{\d h}\Upsilon + \zeta_{\bot}\frac{\d \Upsilon}{\d h}\right)\\
&=& \frac{1}{4}\omega y_0^2 \frac{\d\zeta_{\bot}}{\d h}\left(\Upsilon+\frac{1}{4}Sp\frac{\partial\Upsilon}{\partial Sp}\right)\cdot
\label{active_grad}
\end{eqnarray}
The main difference between Eq.~(\ref{active_grad}) and Eq.~(\ref{grad}) is that the scaling function $\Upsilon$ is a function of both the elastohydrodynamic length scale parametrized by $Sp$ and an active length scale defined by $kL$; the interaction of these two lengths leads to several non-intuitive results, as shown below.  From a dimensional standpoint, we can write 
\begin{equation}
\frac{d}{dh}\langle F \rangle  = \frac{1}{4}\omega y_0^2 \bigg| \frac{\d \zeta_{\bot}}{\d h}\bigg| \Upsilon'
\end{equation}
where, as above, we have defined the dimensionless force gradient as
\begin{equation}
\Upsilon' =- \left(\Upsilon+\frac{1}{4}Sp\frac{\partial\Upsilon}{\partial Sp}\right)\cdot
\end{equation}
Anticipating that the mean force, $\langle F \rangle$, can change sign, we compute the gradient in the norm squared of the force and get
\begin{eqnarray}
\frac{d}{dh}\left[\frac{1}{2}\langle F \rangle^2 \right]&=&
\frac{1}{16}\omega^2 y_0^4 \bigg| \frac{\d \zeta_{\bot}}{\d h}\bigg| 
 \zeta_{\bot} 
 \Upsilon \Upsilon',
\end{eqnarray}
so that $ \Upsilon \Upsilon'$ is the dimensionless gradient in the norm (squared) of the propulsive force.

\begin{figure*}[t!]
\includegraphics[width=.8\textwidth]{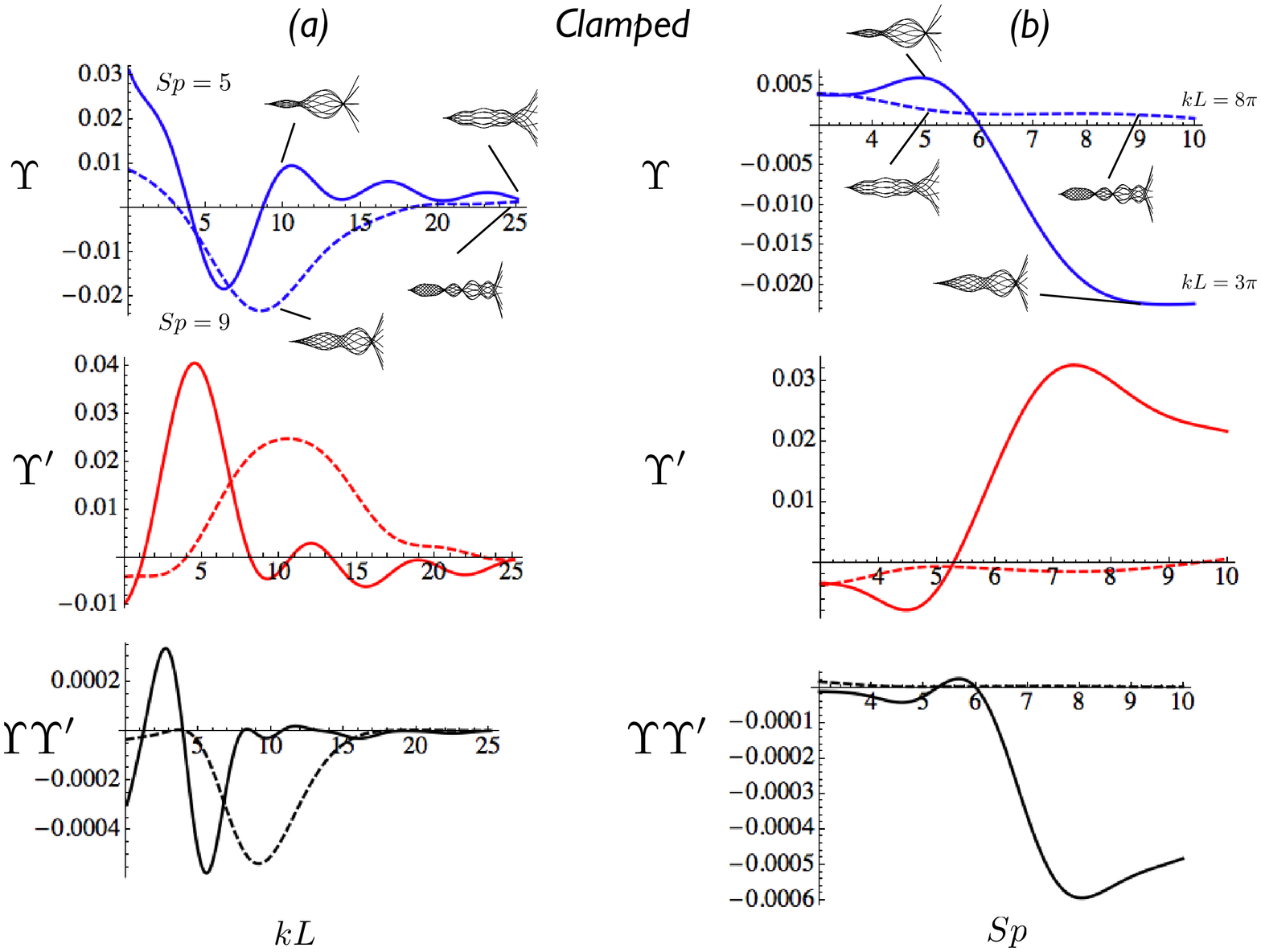}
\caption{\label{forceclamped} (color online) Active filament with prescribed activity and clamped boundary conditions:  dimensionless force, $\Upsilon$, force gradient, $\U'$, and gradient of the square of the norm of the force, $\Upsilon\Upsilon'$, for various values of ($Sp$, $kL$).  The corresponding shapes are displayed at their representative values on the curves.  
(a) $\U$ (blue, top), $\U'$ (red, middle) and $\U\U'$ (bottom, black) as a function of $kL$ for $Sp=5$ (solid) and 9 (dashed);  $kL$ runs from 0.1 to $8\pi$ (b) All three again as a function of $Sp$ for $kL=3\pi$ (solid) and $kL=8\pi$ (dashed); $Sp$ varies from 3 to 10. The axis limits were chosen to cover a wide range of biologically relevant filaments and viscosity solutions, keeping in mind that for small $Sp$ the rigid rod limit renders the model inaccurate and is irrelevant for biological locomotion.}
\end{figure*}

\begin{figure*}[t]
\includegraphics[width=.8\textwidth]{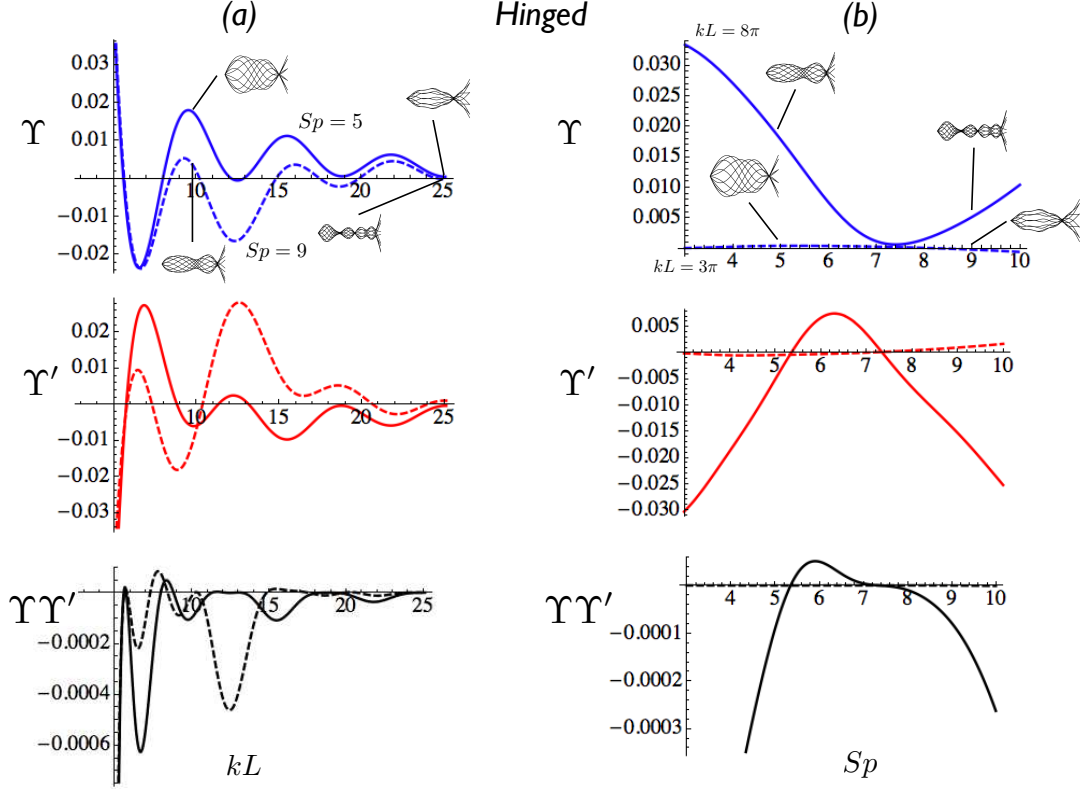}
\caption{\label{forcehinged} (color online) Active filament with prescribed activity and hinged boundary conditions:  Dimensionless force, $\Upsilon$, force gradient, $\Upsilon'$, and gradient of the norm squared of the force, $\U\U'$, for various values of ($Sp$, $kL$).  The corresponding shapes are displayed at their representative values on the curves.  
(a) $\U$ (blue, top), $\U'$ (red, middle) and $\U\U'$ (bottom, black) as a function of $kL$ for $Sp=5$ (solid) and 9 (dashed);   $kL$ varies from 0.1 to $8\pi$ (b) All three again as a function of $Sp$ for $kL=3\pi$ (solid) and $kL=8\pi$ (dashed); here  $Sp$ runs from 3 to 10.}
\end{figure*}

In Fig.~\ref{forceclamped}a we display $\Upsilon$ (top, blue), $\U'$ (middle, red) and $\U\U'$ (bottom, black) for $Sp=5$ (solid line) and $Sp=9$ (dashed line) as a function of $kL$ in the case of clamped boundary conditions.  In Fig.~\ref{forceclamped}b we show the same system, only this time several representative values of $kL$ are chosen and the resulting dimensionless force  and force gradients  are plotted as a function of $Sp$. Similar results are shown in  Fig.~\ref{forcehinged} in the case of  hinged boundary conditions.

First, we observe that although the activity wave is always traveling from the body of the cell to the tip of the active filament ($\omega/k = c > 0$), the propulsive force can change sign: $\Upsilon > 0$ means a force on the fluid in the same direction as the wave, and therefore (if the cell was free to move) swimming in the direction opposite to the wave. As a difference, $\Upsilon <0 $ means the generation of a force on the fluid opposite to the wave propagation, and therefore swimming along the direction of the wave propagation. Recall that in the passive case analyzed in the previous section, we always had a positive force. We also observe here, in general, a non-monotonic variation of the propulsive force with both the activity wavelength (through $kL$) and frequency (through $Sp$). We further note the importance of the boundary conditions as markedly different results are obtained in Fig.~\ref{forceclamped} and Fig.~\ref{forcehinged}.

The second important results to note from Figs.~\ref{forceclamped} and \ref{forcehinged} are the variations in the sign of the force gradient. In all cases where the product 
$\Upsilon \U '> 0$, the presence of the boundary causes the magnitude of the propulsive force to decrease. The opposite is true in all cases where $\Upsilon \U ' < 0$. It  should be obvious from  Figs.~\ref{forceclamped} and \ref{forcehinged}  that the sign of the force gradient displays a complex dependence on the parameters ($Sp$, $kL$), as well as on the type of boundary conditions considered.

To demonstrate the physical significance of our results, we consider as an  example the case of a bull spermatozoa.  For such a cell, we have $L\approx 60$ $\mu$m, $A\approx 10^{-21}$ Nm$^2$, $\omega\approx 10$ Hz and the ``bare" resistive coefficient $(\zeta_\perp)_{\infty}\approx 10^{-3}$ Ns/m$^2$ far from the wall, leading to  $Sp_{\infty}\approx 7$ \cite{Brennen:1977p30}. As discussed above, the wall increases fluid drag, and thus it increases the value of $Sp$.  Since is possible to change the drag coefficient by as much as $50\%$, and  since $Sp\sim \zeta^{1/4}$,  the change in the value of  $Sp$ can be as high as about $10\%$.   The active length scale $kL$ is more difficult to estimate, because the prescribed activity is not immediately obvious through direct observation of the flagellar beat, but  reasonable estimates give $kL$ between $3\pi$ and $5\pi$  \cite{Ishijima1986,Fu:2008p518}. In the context of the prescribed activity model studied here, $kL$ is assumed to not change with distance from the wall.

\begin{figure}[t]
\includegraphics[width=.6\textwidth]{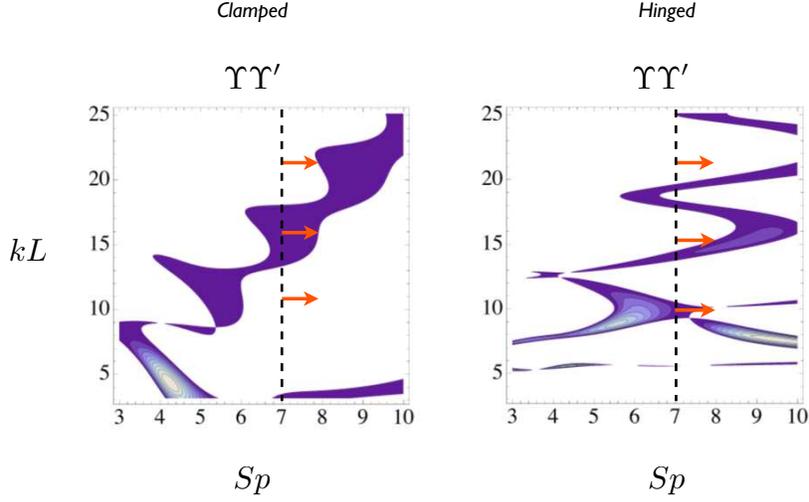}
\caption{\label{contours} (color online) Contour plots for the dimensionless propulsive thrust gradient, $\Upsilon \U '$, as a function of $Sp$ and $kL$. Left: clamped boundary conditions (micro-pipette); right: hinged conditions (optical trap). The force gradient is positive in the filled contour regions; contour lines are $5\times 10^{-3}$ in dimensionless units of force square per unit length.  The arrows indicate the increase in the value of $Sp$ taking place as a beating cell is gradually approached to a solid boundary, starting at $Sp=7$ and for $kL=3\pi$, $5\pi$ and $7\pi$.
}
\end{figure}

We show in Fig.~\ref{contours} contour plots of the gradient of the norm of the propulsive force, $\Upsilon \U '$, as a function of both $Sp$ and $kL$ (left: clamped conditions; right: hinged conditions).  The force gradient is positive in the filled contour regions; contour lines are $5\times 10^{-3}$ in dimensionless units of force square per unit length.   Any given point in the ($Sp$,$kL$) plane gives a particular value of the dimensionless force gradient.  By bringing the beating flagella closer to the wall,  the sperm number is progressively increased, and the value of the new force gradient is found by gradually moving along horizontal lines in Fig.~\ref{contours}  (which are lines of constant  $kL$). We show in  Fig.~\ref{contours}  arrows corresponding to this gradual increase starting at $Sp=7$ and for $kL=3\pi$, $5\pi$ and $7\pi$.  

First we observe, again, that the nature of the boundary condition strongly affects the sign of the force gradient, and experiments performed using optical trapping should give different results from experiments employing micropipettes. Second, we see that three distinct cases are possible depending on the domain crossed by one of the  arrows in Fig.~\ref{contours}. In the first case, the cell away from the wall is in a region where $\Upsilon \U '  $ is negative (white domains in  Fig.~\ref{contours}) and remains in it during the increase of $Sp$; in that case, a measurement would lead to a monotonic increase of the propulsive force as the flagellum comes closer to the boundary. A second case is the one for which $\Upsilon \U ' $ is always positive (for example the middle arrow in the left figure, which remains located inside the positive contour plots), in which case the force would be measured to be monotonically  decreased by the presence of boundaries. Finally, a third situation can arise where the arrow crosses the boundary between a region of positive (respectively  negative) gradient and a region of  negative (respectively positive) gradient, leading to a surprising non-monotonic variation of the propulsive force with the flagellum-wall distance.

%%%%%%%%%%%%%%%%%%%%
%%%%%%%%%%%%%%%%%%%%

\subsection{Self-organized axonemal beating}

The prescribed activity model studied in the previous section  assumes that the internal activity of the flagellum is not modified by the change in the fluid friction near a wall.  In a more physically realistic model, the oscillations of the flagellum would arise from a self-organized motion of the molecular motors, and thus the internal force generation would in turn be a function of the force distribution on the filament (both bending and fluid drag).

We  consider in this section such a model as first introduced by  Camalet and Julicher \cite{Camalet:2000p38}. In this framework, the active force is described by a linear response to the filament sliding, $f=\chi \Delta$. Spontaneous oscillations of flagellum then arise as a self-organized phenomenon for specific values (eigenvalues) of the response function $\chi$. If we use Eq.~(\ref{eom}) with the linear response relationship $f=\chi\Delta\approx a\chi\partial y/\partial x$, then we obtain the following dimensionless eigenvalue equation for the filament amplitude
\begin{gather}
\label{eigenvalueeqn}
\tilde{y}''''-\bar{\chi}\tilde{y}''+iSp^4\tilde{y}=0,
\end{gather}
where $\bar{\chi}=aL^2\chi/A$.  

The amplitude can be solved as $\tilde{y}=\sum_{i=1}^4A_ie^{q_i s}$, where the $q_i$ are the four  solutions to the characteristic equation corresponding to the characteristic modes of Eq.~(\ref{eigenvalueeqn}).  For a given value of  $Sp$, we note that there is an infinite number of discrete eigenvalues $\chi_n$; as in previous work, we will consider only the eigenvalue with lowest norm as they correspond to the lowest degree of motor activity.  
Non-trivial solutions to Eq.~\eqref{eigenvalueeqn} exist only for certain critical pairs of parameters $(Sp_c,\chi_c)$, set by the boundary conditions \cite{Camalet:2000p38,Hilfinger:2009p539}
\begin{gather}
\sum_{i=1}^4M_{ij}A_j=0,
\end{gather}
where the matrix $\textbf{M}$ is defined by the boundary conditions (clamped or hinged, same as in the previous section).  Since the solution, $\tilde y$, is an eigenfunction for a linear equation, it is known only up to a multiplicative constant, and the absolute magnitude for the oscillations can therefore not be obtained. 
Away from a wall, it has been shown theoretically and experimentally that the beating amplitude depends only weakly on nonlinear corrections \cite{Hilfinger:2009p539,RiedelKruse:2007p549}, so we will proceed by writing the (arbitrary)  amplitude of the filament oscillations as $y_0$, which we assume is the magnitude of the first component of the eigenvector (i.e. $A_1=y_0$), and is assumed to remain constant.

The propulsive force acting from the flagellum on the fluid in this case is given by 
\begin{gather}\label{prop_self}
\langle F \rangle =\frac{1}{4}\omega\zeta_{\bot}y_0^2 \Lambda (Sp),
\end{gather}
where
\begin{gather}
\Lambda(Sp)=Im\left[\int_0^1{\tilde{y}^*\frac{\partial \tilde{y}}{\partial x}dx}\right]\cdot
\end{gather}
and which can be written more explicitly as
\begin{gather}\label{complex}
\Lambda(Sp) = Im\left[\int_0^1{\sum_{i=1}^4A_i^*e^{q_i^*x}\sum_{j=1}^4A_jq_je^{q_jx}dx}\right],
\end{gather}
In Eq.~\eqref{complex},  the $A_i$'s and $q_i$'s are all implicit functions of the response $\chi$, which is in turn a function of $Sp$.  These functions are all known, albeit verbose, and thus allow us to explicitly write the  force in terms of $Sp$.

Near a wall, the fluid friction modifies the fluid resistance coefficient, $\zeta_\perp$, but the bending modulus $A$ remains constant. In order to elucidate the variation of the propulsive force with a change in flagellum-wall distance, we thus need to know how both the beat frequency, $\omega$, and the response function, $\chi$, vary. Without further biological information about the behavior of molecular motors under changing load, we now have to make modeling assumptions.

{The functional dependence of the response function $\chi$,  on the oscillation frequency and other materiel parameters is, in general, unknown.  In order to satisfactorily solve Eq.~\eqref{eigenvalueeqn} we need to find the complex eigenvalues that allow for non-trivial solutions of the equation to exist. To fully model the active system, the filament response function should be derived from a model for the molecular motors, and without such an explicit model, $\chi$ could very generally be a function of on $Sp$, ATP production  and concentration (i.e. activity level), load distribution, structural inhomogeneities, and  any other parameter(s) that govern activity in the axoneme.  The most general solution would thus require a detailed model of the molecular motors  (see e.g. \cite{Camalet:2000p38,Hilfinger:2009p539,RiedelKruse:2007p549}).}

{Since our primary focus in this paper is to explore the hydrodynamic consequences of activity, boundaries, and elasticity, we will make below several modeling assumptions in order to examine extreme cases, by essentially specifying the functional dependence of $\chi$. In general both the frequency $\omega$ and the response function $\chi$ will vary, but we are going to assume  here that one of them  remains essentially constant as the wall-flagellum distance is varied. We thus assume that one parameter  shows a strong variation with $h$ whereas the other depends only weakly on $h$.}

In the first case,  the linear response function remains constant, such that the only way for the eigensolution to Eq.~\eqref{eigenvalueeqn} to have non-trivial values requires the  oscillation frequency, $\omega$, to change in such a manner that $Sp$ remains constant. Given the definition of $Sp$, this means that one would observe experimentally a frequency change given by ${\omega({h})}/{\omega (h={\infty})}={\zeta_{\perp}(h= \infty)}/{\zeta_\perp (h)}$, but no waveform variation. In that case, the only change in the propulsive force, Eq.~\eqref{prop_self},  would arise from the variation of $\zeta_\perp$ with $h$, and therefore one would experimentally  measure an increase in the magnitude of the force  near boundaries.

The second, more complex, case is one in which the frequency of oscillation of the filament would remain fixed.  {Although there is no experimental evidence that demonstrates that $\omega$ does remain constant as the distance between the cell and the boundary is changing, it is a reasonable assumption to make in the absence of this information.  This assumption, however, puts a stringent constraint on the response function $\chi$; in order to provide non-trivial solutions to Eq.~\eqref{eigenvalueeqn} the response function must be only a function of $Sp$ in such a fashion as to match the eigenvalues exactly.} In this case, varying the distance between the flagellum and the wall, $h$, modifies the value of $Sp$. As a result of changes in $Sp$, both the response function $\chi$, and the eigenfunction,  will be continuously modified, which will produce a non-trivial variation in the propulsive force. 
As in the previous section, we will write the force gradient formally as
\begin{equation}
\frac{\d}{\d h} \langle F \rangle = \frac{1}{4}\omega y_0^2 \bigg|\frac{\d\zeta_{\bot}}{\d h}\bigg|  \Lambda '.
\end{equation}

\begin{figure}[t]
\includegraphics[width=.7\textwidth]{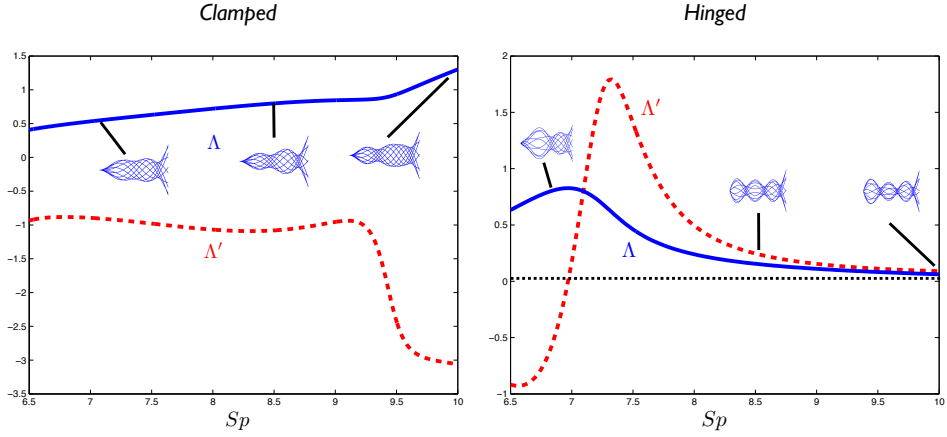}
\caption{\label{forceresponse} (color online) 
Dimensionless force ($\Lambda$, blue line) and force gradient  ($\Lambda'$, red dashed line) for the linear response model with clamped  (left) and hinged (right) boundary conditions, as a function of $Sp$.  Characteristic flagellar beat patterns are also shown at several values of $Sp$.}
\end{figure}

For this scenario, we plot in Fig.~\ref{forceresponse} the dimensionless propulsive force ($\Lambda $) and force gradient ($\Lambda'$) for the clamped and hinged boundary conditions respectively, as well as  several exemplary beat patterns.  With this modeling approach, we see that  $\Lambda $ is always positive. 
Using the notation of Fig.~\ref{schematic}, the flagellum is therefore always pushing the fluid along the $x$ direction, and is thus expected to swim in the $-x$ direction.
Furthermore, we obtain that the sign the force gradient depends on the nature of the boundary conditions. 
In the clamped case, we get that $\Lambda '$ is always negative, and therefore the presence of boundaries systematically increases the propulsive force. In contrast, in the hinged case, the function  $\Lambda '$ is seen to be negative for $Sp $ below 7, and positive otherwise.  In that case, and similarly to what was reported experimentally using optical trapping in  Ref.~\cite{Lauga10}, the measurements would show a decrease in the propulsive force near boundaries.

\section{discussion}
\label{discussion}

Biological cells do not swim in a vacuum; the environment itself is what makes swimming possible, and thus we must consider characteristics of the surroundings that may modify motility behavior.  Since flagellated organisms generate propulsion by actuating an elastic filament to do work against their viscous environment, the specific response of that environment can be crucial to understanding the overall locomotion characteristics.   The particular environmental effect that we have studied in this paper is the modification of fluid drag by the presence of a no-slip boundary, and the balance between the deformable waveform of the flagellum and the viscous fluid forces that generate propulsive thrust.  

For driven filaments, we have shown that fluid drag plays a dual role: not only does it change the propulsion generated by a given filament waveform, but it also affects the waveform itself expressed by the beating filaments. For passively actuated filaments, the resulting wall effect is a systematic  increase of the propulsive force that the beating filament imparts on the surrounding fluid in the case of displacement-driven actuation, while a decrease is obtained in the case of force-driven actuation.  In contrast, for active filaments as models for eukaryotic flagella, the modification to the propulsive force depends sensitively on a combination of the flagellar material properties, the boundary condition applied to the flagellum,  and the manner in which the molecular motors organize to cause oscillation; different values of parameters can increase, decrease or even display non-monotonic influence on the cellular propulsive force. 

Using simple scaling arguments, let us finally estimate the expected size of the propulsive force change  induced by  a wall on an active flagella . Let us  compare the order of magnitude for the force far away from the boundary to the average change in force between near- and far-field.  Far from the wall the propulsive force scales as $F_\infty   \sim  \omega y_0^2 (\zeta_\perp)_{\infty} \Upsilon $.  Since our calculation for the force change due to a wall focuses on  the near field, we can estimate the average change in force as $ \Delta F\sim L \times d  F  / dh $ as $L$, the cell length, gives approximately the spatial range over which the near-field matches with the far-field.   Using the estimate ${d}F/{dh}    \sim-  \omega y_0^2 ({\d \zeta_{\bot}}/{\d h})  \Upsilon'$ we therefore get
\begin{equation}
\frac{\Delta F}{F_\infty}\sim - \frac{  L({\d \zeta_{\bot}}/{\d h})  \Upsilon' }{ (\zeta_\perp)_{\infty} \Upsilon }\cdot
\end{equation}
According to Eq.~\eqref{zeta_wall}, we have ${\d\zeta_{\bot}}/{\d h}\sim-{\zeta_{\bot}^2}/{h\mu}$ hence the scaling becomes
\begin{equation}
\frac{\Delta F}{F_\infty}
\sim \frac{  L {\zeta_{\bot}^2}   }{ h\mu (\zeta_\perp)_{\infty} }
\left(\frac{\Upsilon'}{\Upsilon }\right)\cdot
\end{equation}
Since we know that  $(\zeta_\perp)_{\infty}\approx \mu/\ln(2L/a)$ and $\zeta_\perp\sim \mu/\ln(2h/a)$, we get the final scaling relationship
\begin{equation}
\frac{\Delta F}{F_\infty}
\sim  \frac{\ln(2L/a) }{  [\ln(2h/a)]^2 }
\left(\frac{\Upsilon'}{\Upsilon }\right)\frac{L}{h}\cdot
\end{equation}
From the results  obtained  above, we observe that ${\Upsilon'}/{\Upsilon }\sim \pm 1$. 
For human spermatozoa,  the parameters are $L\approx 40$ $\mu$m, $a\approx 0.20$ $\mu$m. In the experiment of Ref.~\cite{Lauga10} the near-field measurements get as close as $h=5$ $ \mu m$  which leads to ${\Delta F}/{F_\infty}\sim \pm 1$. This  simple order-of-magnitude calculation shows that the force could be expected to be  changed by order one by the introduction of a boundary. In the experiment conducted in Ref.~\cite{Lauga10} the force was measured to be reduced by a factor of three, which is consistent with this simple estimate.

\section*{Acknowledgments}
We thank Michael Berns and Linda Shi for useful discussions, and for the collaboration which planted the seeds for this theoretical investigation.  Funding by the National Science Foundation (grant CBET-0746285 to EL) is gratefully acknowledged. 

\bibliography{flagellar_propulsion}

\end{document}